\documentclass{siamltex}

\def\off{{OFF}}
\def\ouralg{{OP}}

\begin{document}

\vspace*{-20pt}
\slugger{sidma}{1998}{0}{0}{000--000}

\title{On-Line Difference Maximization\thanks{Received by the editors
July 29, 1996; accepted for publication (in revised form) March 5,
1998; published electronically DATE.}}
\author{Ming-Yang Kao\thanks{Department of Computer Science, Yale
University, New Haven, CT\ \ 06520 ({\tt kao-ming-yang@cs.yale.edu}).
Supported in part by NSF Grant CCR-9531028.}
\and
Stephen R. Tate\thanks{Department of Computer Science, University of
North Texas, Denton, TX\ \ 76208 ({\tt srt@cs.unt.edu}).  Supported in
part by NSF Grant CCR-9409945.}}
\maketitle

\begin{abstract}
In this paper we examine problems motivated by on-line financial
problems and stochastic games.  In particular, we consider a sequence
of entirely arbitrary distinct values arriving in random order, and
must devise strategies for selecting low values followed by high
values in such a way as to maximize the expected gain in rank from low
values to high values.

First, we consider a scenario in which only one low value and one high
value may be selected.  We give an optimal on-line algorithm for this
scenario, and analyze it to show that, surprisingly, the expected gain
is $n-O(1)$, and so differs from the best possible {\em off-line} gain
by only a constant additive term (which is, in fact, fairly small --- at
most 15).

In a second scenario, we allow multiple nonoverlapping low/high
selections, where the total gain for our algorithm is the sum of the
individual pair gains.  We also give an optimal on-line algorithm for
this problem, where the expected gain is $n^2/8-\Theta(n\log
n)$.  An analysis shows that the optimal expected off-line gain is
$n^2/6+\Theta(1)$, so the performance of our on-line
algorithm is within a factor of $3/4$ of the best off-line
strategy.
\end{abstract}



\begin{PII}
S0895480196307445
\end{PII}

\pagestyle{myheadings}
\thispagestyle{plain}
\markboth{M.~Y. KAO AND S.~R. TATE}{ON-LINE DIFFERENCE MAXIMIZATION}
 
\section{Introduction}
In this paper, we examine the problem of accepting values from an
on-line source and selecting values in such a way as to maximize the
difference in the ranks of the selected values.
The input values can be arbitrary distinct real numbers, and thus we
cannot determine with certainty the actual ranks of any input values
until we see all of them.  Since we only care about their ranks, an
equivalent way of defining the input is as a sequence of $n$ integers
$x_1,x_2,\ldots,x_n$, where $1\leq x_i\leq i$ for all
$i\in\{1,\ldots,n\}$, and input $x_i$ denotes the rank of the $i$th
input item among the first $i$ items.  These ranks uniquely define an
ordering of all $n$ inputs, which can be specified with a sequence of
ranks $r_1,r_2,\ldots,r_n$, where these ranks form a permutation of
the set $\{1,2,\ldots,n\}$.  We refer to the $r_i$ ranks as
{\em final ranks}, since they represent the rank of each item among the
final set of $n$ inputs.  We assume that the inputs come from a
probabilistic source such that all permutations of $n$ final ranks are
equally likely.

The original motivation for this problem came from considering on-line
financial problems~\cite{AS89,CCEKL95,Cover84,EFKT92,Fama76}, where
maximizing the difference between selected items naturally corresponds
to maximizing the difference between the buying and selling prices of
an investment.  While we use generic terminology in order to
generalize the setting (for example, we make a ``low selection''
rather than pick a ``buying price''), many of the problems examined in
this paper are easily understood using notions from investing.  This
paper is a first step in applying on-line algorithmic techniques to
realistic on-line investment problems.

While the original motivation comes from financial problems, the
current input model has little to do with realistic financial markets,
and is selected for its mathematical cleanness and its relation to
fundamental problems in stochastic games.  The main difference between
our model and more realistic financial problems is that in usual stock
trading, optimizating rank-related quantities is not always correlated
to optimizing profits in the dollar amount.  However, there are some
strong similarities as well, such as exotic financial derivatives
based on quantities similar to ranks~\cite{WilmottHD}.

The current formulation is closely related to an important
mathematical problem known as the {\em secretary
problem}~\cite{gardner:60,cmrs:64}, which has become a standard
textbook example~\cite{billingsley:86,crs:71,whittle:82}, and has been
the basis for many interesting extensions
(including~\cite{amw:95,mucci:73,rp:76,sd:75,tamaki:79}).  The
secretary problem comes from the following scenario: A set of
candidates for a single secretarial position are presented in random
order.  The interviewer sees the candidates one at a time, and must
make a decision to hire or not to hire immediately upon seeing each
candidate.  Once a candidate is passed over, the interviewer may not
go back and hire that candidate.  The general goal is to maximize
either the probability of selecting the top candidate, or the expected
rank of the selected candidate.  This problem has also been stated
with the slightly different story of a princess selecting a
suitor~\cite[p. 110]{billingsley:86}.  More will be made of the
relationship between our current problem and the secretary problem in
\S\ref{sec:single}, and for further reading on the secretary
problem, we refer the reader to the survey by
Freeman~\cite{freeman:83}.

As mentioned above, we assume that the input comes from a random
source in which all permutations of final ranks $1,2,\ldots,n$ are
equally likely.  Thus, each rank $x_i$ is uniformly distributed over
the set $\{1,2,\ldots,i\}$, and all ranks are independent of one
another.  In fact, this closely parallels the most popular algorithm
for generating a random permutation~\cite[p. 139]{knuth:81b}.
A natural question to ask is, knowing the relative rank $x_i$ of the
current input, what is the expected final rank of this item (i.e.,
$E[r_i|x_i]$)?  Due to the uniform nature of the input source, the
final rank of the $i$th item simply scales up with the number of items
left in the input sequence, and so $E[r_i|x_i]=\frac{n+1}{i+1}x_i$ (a
simple proof of this is given in Appendix A).

Since all input ranks $x_i$ are independent and uniformly distributed,
little can be inferred about the future inputs.  We consider games in
which a player watches the stream of inputs, and can select items as
they are seen; however, if an item is passed up then it is gone for
good and may not be selected later.  We are interested in strategies
for two such games:
\begin{itemize}
\item Single pair selection: In this game, the player
should make two selections, the first being the {\em low selection}
and the second being the {\em high selection}.  The goal of the player
is to maximize the difference between the final ranks of these two
selections.  If the player picks the low selection upon seeing input
$x_\ell$ at time step $\ell$, and picks the high selection as input
$x_h$ at time step $h$, then the {\em profit} given to the player at
the end of the game is the difference in final ranks of these items:
$r_h-r_\ell$.
\item Multiple pair selection: In this game, the player
makes multiple choices of low/high pairs.  At the end of the game the
difference in final ranks of each selected pair of items is taken, and
the differences for all pairs are added up to produce the player's
final profit.
\end{itemize}
The strategies for these games share a common difficulty: If the player
waits too long to make the low selection, he risks not having enough
choices for a good high selection; however, making the low selection
too early may result in an item selected before any truly low
items have been seen.  The player in the second game can afford to be
less selective. If one chosen pair does not give a large difference,
there may still be many other pairs that are good enough to make up
for this pair's small difference.

We present optimal solutions to both of the games.  For the first
game, where the player makes a single low selection and a single high
selection, our strategy has expected profit $n-O(1)$. From the
derivation of our strategy, it will be clear that the strategy is
optimal.  Even with full knowledge of the final ranks of all input
items, the best expected profit in this game is less than $n$, and so
in standard terms of on-line performance
measurement~\cite{kmrs:86,st:85}, the competitive
ratio\footnote{``Competitive ratio'' usually refers to the worst-case
ratio of on-line to off-line cost; however, in our case inputs are
entirely probabilistic, so our ``competitive ratio'' refers to
expected on-line to expected off-line cost --- a worst-case measure
doesn't even make sense here.} of our strategy is one.  The strength
of our on-line strategy is rather intriguing.

For the second game, where multiple low/high pairs are selected, we
provide an optimal strategy with expected profit
$\frac{1}{8}n^2-O(n\log n)$.  For this problem, the optimal off-line
strategy has expected profit of approximately $\frac{1}{6}n^2$,
and so the competitive ratio of our strategy is $\frac{4}{3}$.

\section{Single Low/High Selection}
\label{sec:single}
This section considers a scenario in which the player may pick a
single item as the low selection, and a single later item as the high
selection.  If the low selection is made at time step $\ell$ and the
high selection is made at time step $h$, then the expected profit is
$E[r_h-r_{\ell}]$.  The player's goal is to use a strategy for picking
$\ell$ and $h$ in order to maximize this expected profit.

As mentioned in the previous section, this problem is closely related
to the secretary problem.  A great deal of work has been done on the
secretary problem and its variations, and this problem has taken a
fundamental role in the study of games against a stochastic opponent.
Our work extends the secretary problem, and gives complete solutions
to two natural variants that have not previously appeared in the
literature.

Much insight can be gained by looking at the optimal solution to the
secretary problem, so we first sketch that solution below (using
terminology from our problem about a ``high selection'').  To maximize
the expected rank of a single high selection, we define the optimal
strategy recursively using the following two functions:

\vspace*{1em}

\begin{tabular}{cl}
${\cal H}_n(i)$: & This is a limit such that the player selects the
current item if\\ & $x_i\geq {\cal H}_n(i)$. \\
$R_n(i)$: & This is the expected final rank of the high selection if
the optimal\\ & strategy is followed starting at the $i$th time step.
\end{tabular}

\vspace*{1em}

Since all permutations of the final ranks are equally likely, if the
$i$th input item has rank $x_i$ among the first $i$ data items, then
its expected final rank is $\frac{n+1}{i+1}x_i$.  Thus, an optimal
strategy for the secretary problem is to select the $i$th input item
if and only if its expected final rank is better than could be
obtained by passing over this item and using the optimal strategy from
step $i+1$ on.  In other words, select the item at time step $i<n$ if
and only if
\[  \frac{n+1}{i+1}x_i \geq R_n(i+1) . \]
If we have not made a selection before the $n$th step, then we must
select the last item, whose rank is uniformly distributed over the
range of integers from 1 to $n$ --- and so the expected final rank in
that case is $R_n(n)=\frac{n+1}{2}$.  For $i<n$ we can also define
\[  {\cal H}_n(i) = \left\lceil \frac{i+1}{n+1} R_n(i+1) \right\rceil
, \]
and to force selection at the last time step define ${\cal H}_n(n)=0$.
Furthermore, given this definition for ${\cal H}_n(i)$, the optimal
strategy at step $i$ depends only on the rank of the current item
(which is uniformly distributed over the range $1,\ldots,i$) and the
optimal strategy at time $i+1$.  This allows us to recursively define
$R_n(i)$ as follows when $i<n$:
\begin{eqnarray*}
 R_n(i) & = & \frac{{\cal H}_n(i)-1}{i} R_n(i+1) +
              \sum_{j={\cal H}_n(i)}^i \frac{1}{i}\cdot \frac{n+1}{i+1} j \\
 & = & \frac{{\cal H}_n(i)-1}{i} R_n(i+1) +
       \frac{n+1}{i(i+1)}\cdot \frac{(i+{\cal H}_n(i))(i-{\cal H}_n(i)+1)}{2} \\
 & = & \frac{{\cal H}_n(i)-1}{i}
         \left( R_n(i+1) - \frac{n+1}{2(i+1)} {\cal H}_n(i) \right) +
         \frac{n+1}{2} .
\end{eqnarray*}
Since ${\cal H}_n(n)=0$ and $R_n(n)=\frac{n+1}{2}$, we have a full
recursive specification of both the optimal strategy and the
performance of the optimal strategy.  The performance of the optimal
strategy, taken from the beginning, is $R_n(1)$.  This value can be
computed by the recursive equations, and was proved by Chow et
al. to tend to $n+1-c$, for $c\approx 3.8695$, as
$n\rightarrow\infty$~\cite{cmrs:64}.  Furthermore, the performance
approaches this limit from above, so for all $n$ we have performance
greater than $n-2.87$.

For single pair selection, once a low selection is made we want to
maximize the expected final rank of the high selection.  If we made
the low selection at step $i$, then we can optimally make the high
selection by following the above strategy for the secretary problem,
which results in an expected high selection rank of $R_n(i+1)$.
How do we make the low selection?  We can do this
optimally by extending the recursive definitions given above with two
new functions:

\vspace*{1em}

\begin{tabular}{cl}
${\cal L}_n(i)$: & This is a limit such that the player selects the
current item if\\ & $x_i\leq {\cal L}_n(i)$. \\
$P_n(i)$: & This is the expected high-low difference if the optimal
strategy for\\ & making the low and high selections is followed 
starting at step $i$.
\end{tabular}

\vspace*{1em}

Thus, if we choose the $i$th input as the low selection, the expected
profit is $R_n(i+1)-\frac{n+1}{i+1}x_i$. We should select this
item if that expected profit is no less than the expected profit if we
skip this item.  This leads to the definition of ${\cal L}_n(i)$:
\[  {\cal L}_n(i) = \left\{
   \begin{array}{ll}
        0 & \mbox{ if $i=n$ , } \\
        \left\lfloor \frac{i+1}{n+1}\left( R_n(i+1)-P_n(i+1) 
             \right)\right\rfloor & \mbox{ if $i<n$ .}
   \end{array}\right.
\]
Using ${\cal L}_n(i)$, we derive the following profit function:
\[  P_n(i) = \left\{
   \begin{array}{ll}
        0 & \mbox{ if $i=n$ , } \\
        P_n(i+1) + \frac{{\cal L}_n(i)}{i}\left(
          R_n(i+1) - P_n(i+1) - \frac{n+1}{i+1}\cdot 
          \frac{{\cal L}_n(i)+1}{2}\right)
          & \mbox{ if $i<n$ . }
   \end{array}\right.
\] From the derivation, it is clear that this is the optimal strategy,
and can be implemented by using the recursive formulas to compute the
${\cal L}_n(i)$ values.  The expected profit of our algorithm is given
by $P_n(1)$, which is bounded in the following theorem.

\begin{theorem}
Our on-line algorithm for single low/high selection is optimal and has
expected profit $n-O(1)$.
\end{theorem}

\begin{proof}
It suffices to prove that a certain inferior algorithm has
expected profit $n-O(1)$.  The inferior algorithm is as follows: Use
the solution to the secretary problem to select, from the first
$\lfloor n/2\rfloor$ input items, an item with the minimum expected
final rank.  Similarly, pick an item with maximum expected rank from
the second $\lceil n/2\rceil$ inputs.  For simplicity, we initially
assume that $n$ is even; see comments at the end of the proof for
odd $n$.  Let $\ell$ be the time step in which the low selection is
made, and $h$ the time step in which the high selection is made.
Using the bounds from Chow et al.~\cite{cmrs:64}, we can
bound the expected profit of this inferior algorithm by
\begin{eqnarray*}
 E[r_h-r_{\ell}] & = & E[r_h] - E[r_l]
    \geq \frac{n+1}{n/2+1}(n/2+1-c) - \frac{n+1}{n/2+1} c \\
    & = & \frac{n+1}{n+2}(n+2-4c)
    = n+1-4c+\frac{4c}{n+2}\ .
\end{eqnarray*}
Chow et al.~\cite{cmrs:64} show that $c\leq 3.87$, and so the
expected profit of the inferior algorithm is at least $n-14.48$.
For odd $n$, the derivation is almost identical, with only a change in
the least significant term;  specifically, the expected profit of the
inferior algorithm for odd $n$ is $n+1-4c+\frac{4c}{n+3}$, which again
is at least $n-14.48$.
\qquad\end{proof}

\section{Multiple Low/High Selection}
This section considers a scenario in which the player again selects a
low item followed by a high item, but may repeat this process as often
as desired.  If the player makes $k$ low and high selections at time
steps $\ell_1, \ell_2,\ldots,\ell_k$ and $h_1, h_2, \ldots, h_k$,
respectively, then we require that
\[ 1 \leq \ell_1 < h_1 < \ell_2 < h_2 < \cdots < \ell_k < h_k \leq n . \]
The expected profit resulting from these selections
is 
\[  E[r_{h_1}-r_{\ell_1}] + E[r_{h_2}-r_{\ell_2}] + \cdots +
           E[r_{h_k}-r_{\ell_k}] . \]

\subsection{Off-line Analysis}
\label{sec:offline}

Let {\em interval $j$} refer to the time period between the
instant of input item $j$ arriving and the instant of input item $j+1$
arriving.  For a particular sequence of low and high selections, we
call interval $j$ {\em active} if $\ell_i\leq j<h_i$ for some
index $i$.  We then amortize the total profit of a particular
algorithm $B$ by defining the amortized profit $A_B(j)$ for interval
$j$ to be
\[  A_B(j) = \left\{ \begin{array}{ll}
          r_{j+1}-r_j & \mbox{ if interval $j$ is active,} \\
          0 & \mbox{ otherwise.} \end{array}\right. \]
Note that for a fixed sequence of low/high selections,
the sum of all amortized profits is exactly the total profit, i.e.,
\[  \sum_{j=1}^n A_B(j) = 
       \sum_{j=\ell_1}^{h_1-1} (r_{j+1}-r_j) +
       \sum_{j=\ell_2}^{h_2-1} (r_{j+1}-r_j) + \cdots +
       \sum_{j=\ell_k}^{h_k-1} (r_{j+1}-r_j) \]
\[  = (r_{h_1}-r_{\ell_1}) + (r_{h_2}-r_{\ell_2}) + \cdots +
      (r_{h_k}-r_{\ell_k}) . \]

For an off-line algorithm to maximize the total profit we need to
maximize the amortized profit, which is done for a particular sequence
of $r_i$'s by making interval $j$ active if and only if $r_{j+1}>r_j$.
Translating this back to the original problem of making low and
high selections, this is equivalent to identifying all maximal-length
increasing intervals and selecting the beginning and ending points of
these intervals as low and high selections, respectively.  These
observations and some analysis give the following lemma.

\begin{lemma}
The optimal off-line algorithm just described has expected profit
$\frac{1}{6}\left(n^2-1\right)$.
\end{lemma}

\begin{proof}
This analysis is performed by examining the expected amortized profits
for individual intervals.  In particular, for any interval $j$,
\begin{eqnarray*}
 E[A_\off(j)] & = & Pr[r_{j+1}>r_j]\cdot E[A_j|r_{j+1}>r_j] +
              Pr[r_{j+1}<r_j]\cdot E[A_j|r_{j+1}<r_j] \\
        & = & \frac{1}{2}\cdot E[r_{j+1}-r_{j}|r_{j+1}>r_j] +
              \frac{1}{2}\cdot 0 \\
        & = & \frac{1}{2} \sum_{i=1}^{n-1} 
         \sum_{k=i+1}^{n} 
\frac{Pr[r_{j+1}=k\mbox{ and }r_j=i]}{Pr[r_{j+1}>r_j]}\cdot (k-i) \\
        & = & \frac{1}{2} \sum_{i=1}^{n-1} 
         \sum_{k=i+1}^{n} \frac{2}{n(n-1)} (k-i) \\
        & = & \frac{1}{2}\cdot \frac{2}{n(n-1)}\cdot \frac{(n+1)n(n-1)}{6} \\
        & = & \frac{n+1}{6} .
\end{eqnarray*}
Since there are $n-1$ intervals and the above analysis is independent
of the interval number $j$, summing the amortized profit over all
intervals gives the expected profit stated in the lemma.
\qquad\end{proof}

\subsection{On-line Analysis}
\label{sec:multiple}

In our on-line algorithm for multiple pair selection, there are two
possible states: {\sc free} and {\sc holding}.  In the {\sc free}
state, we choose the current item as a low selection if $x_i <
\frac{i+1}{2}$; furthermore, if we select an item then we move from
the {\sc free} state into the {\sc holding} state.  On the other hand,
in the {\sc holding} state if the current item has $x_i >
\frac{i+1}{2}$, then we choose this item as a high selection and move into
the {\sc free} state.  We name this algorithm \ouralg, which can stand
for ``opportunistic'' since this algorithm makes a low selection
whenever the probability is greater than $\frac{1}{2}$ that the next
input item will be greater than this one.  Later we will see that the
name \ouralg\ could just as well stand for ``optimal,'' since this
algorithm is indeed optimal.

The following lemma gives the expected profit of this algorithm.
In the proof of this lemma we use the following equality:
\[ \sum_{i=1}^k \frac{2i}{2i+1} = k+1+\frac{1}{2}H_k-H_{2k+1} . \]

\begin{lemma}
\label{lem:multialg}
The expected profit from our on-line algorithm is
\[  E[P_{\ouralg}] = \left\{ \begin{array}{ll}
\displaystyle
   \frac{n+1}{8}\left( n+H_{\frac{n-2}{2}}-2H_{n-1} \right) &
        \mbox{ if $n$ is even,} \\
\ \\
\displaystyle
   \frac{n+1}{8}\left( n+H_{\frac{n-1}{2}}-2H_n + \frac{1}{n} \right) &
        \mbox{ if $n$ is odd.} \\
   \end{array}\right. \]
In cleaner forms we have
$E[P_{\ouralg}]=\frac{n+1}{8}(n-H_n+\Theta(1))=\frac{1}{8}n^2-
\Theta(n\log n)$.
\end{lemma}

\begin{proof}
Let $R_i$ be the random variable of the final rank of the
$i$th input item.  Let $A_\ouralg(i)$ be the amortized cost for
interval $i$ as defined in \S\ref{sec:offline}.  Since
$A_\ouralg(i)$ is nonzero only when interval $i$ is active, 
\begin{eqnarray*}
 E[A_\ouralg(i)] & = &
 E[A_\ouralg(i)|\mbox{Interval $i$ is active}]\cdot Prob[\mbox{Interval $i$ is
active}] \\
 & = & E[R_{i+1}-R_i|\mbox{Interval $i$ is active}]\cdot 
       Prob[\mbox{Interval $i$ is active}] .
\end{eqnarray*}
Therefore,
\begin{eqnarray*}
E[P_\ouralg] & = & \sum_{i=1}^{n-1} E[A_\ouralg(i)] \\
 & = & \sum_{i=1}^{n-1}E[R_{i+1}-R_i|\mbox{Interval $i$ is active}]\cdot 
       Prob[\mbox{Interval $i$ is active}] .
\end{eqnarray*}

Under what conditions is an interval active?  If $x_i<\frac{i+1}{2}$
this interval is certainly active.  If the algorithm was not in the
{\sc holding} state prior to this step, it would be after seeing input
$x_i$.  Similarly, if $x_i>\frac{i+1}{2}$ the algorithm must be in the
{\sc free} state during this interval, and so the interval is not
active.  Finally, if $x_i=\frac{i+1}{2}$ the state remains what it has
been for interval $i-1$.  Furthermore, since $i$ must be odd for this
case to be possible, $i-1$ is even, and $x_{i-1}$ cannot be
$\frac{i}{2}$ (and thus $x_{i-1}$ unambiguously indicates whether
interval $i$ is active).  In summary, determining whether interval $i$
is active requires looking at only $x_i$ and occasionally 
$x_{i-1}$.  Since the expected amortized profit of step $i$ depends on
whether $i$ is odd or even, we break the analysis up into these two
cases below.

\begin{description}
\item[Case 1: $i$ is even.]  Note that
$Prob[x_i<\frac{i+1}{2}]=\frac{1}{2}$, and $x_i$ cannot be exactly
$\frac{i+1}{2}$, which means that with probability $\frac{1}{2}$
interval $i$ is active.  Furthermore, $R_{i+1}$ is
independent of whether interval $i$ is active or not, and so
\begin{eqnarray*}
    E[A_\ouralg(i)|\mbox{Interval $i$ is active}] & = &
    E[R_{i+1}]-E[R_i|\mbox{Interval $i$ is active}] \\
 & = & \frac{n+1}{2} - \frac{n+1}{i+1}\sum_{j=1}^{i/2} \frac{2}{i} j \\
 & = & \frac{n+1}{2} - \frac{n+1}{i+1}\cdot\frac{2}{i}\cdot\frac{i(i+2)}{8} \\
 & = & \frac{n+1}{4}\cdot\frac{i}{i+1} .
\end{eqnarray*}

\item[Case 2: $i$ is odd.]  Since interval 1 cannot be active, we
assume that $i\geq 3$.  We need to consider the case in which
$x_i=\frac{i+1}{2}$, and so
\begin{eqnarray*}
   \hbox to 0in{$Prob[\mbox{Interval $i$ is active}]$\hss} &&\\
   & = & Prob[x_i<\frac{i+1}{2}] + Prob[x_i=\frac{i+1}{2}]\cdot 
              Prob[x_{i-1}<\frac{i}{2}] \\
   & = & \frac{i-1}{2i} + \frac{1}{i}\cdot\frac{1}{2} = \frac{1}{2} .
\end{eqnarray*}
Computing the expected amortized cost of interval $i$ is slightly
more complex than in Case 1.
\begin{eqnarray*}
  \hbox to 0in{$E[A_\ouralg(i)|\mbox{Interval $i$ is active}]$\hss} &&\\
 & = & E[R_{i+1}]-E[R_i|\mbox{Interval $i$ is active}] \\
 & = & \frac{n+1}{2} - 
       \frac{n+1}{i+1}\left(\sum_{j=1}^{(i-1)/2} \frac{2}{i} j +
       \frac{1}{i}\cdot\frac{i+1}{2}\right) \\
 & = & \frac{n+1}{2} - 
       \frac{n+1}{i+1}\left(\frac{2}{i}\cdot\frac{(i-1)(i+1)}{8} +
       \frac{1}{i}\cdot\frac{i+1}{2}\right) \\
 & = & \frac{n+1}{2} - \frac{n+1}{i+1}\cdot\frac{(i+1)(i+1)}{4i} \\
 & = & \frac{n+1}{4}\cdot\frac{i-1}{i} .
\end{eqnarray*}

\end{description}

\noindent
Combining both cases,
\begin{eqnarray*}
  E[P_\ouralg] & = & \sum_{i=1}^{n-1} 
  E[A_\ouralg(i)|\mbox{Interval $i$ is active}]\cdot
     Prob[\mbox{Interval $i$ is active}] \\
  & = & \frac{n+1}{8}\left(
         \sum_{k=1}^{\lfloor(n-2)/2\rfloor} \frac{2k}{2k+1} +
         \sum_{k=1}^{\lfloor(n-1)/2\rfloor} \frac{2k}{2k+1}\right) ,
\end{eqnarray*}
where the first sum accounts for the odd terms of the original sum, and
the second sum accounts for the even terms.

When $n$ is even this sum becomes
\begin{eqnarray*}
E[P_\ouralg] & = & \frac{n+1}{8}\left(
         \sum_{k=1}^{\lfloor(n-2)/2\rfloor} \frac{2k}{2k+1} +
         \sum_{k=1}^{\lfloor(n-1)/2\rfloor} \frac{2k}{2k+1}\right) \\
 & = & \frac{n+1}{8} \left( 2\sum_{k=1}^{(n-2)/2} \frac{2k}{2k+1}\right) \\
 & = & \frac{n+1}{8}\left(n+H_{\frac{n-2}{2}}-2H_{n-1}\right) ,
\end{eqnarray*}
which agrees with the claim in the lemma.  When $n$ is odd the sum can
be simplified as
\begin{eqnarray*}
E[P_\ouralg] & = & \frac{n+1}{8}\left(
         \sum_{k=1}^{\lfloor(n-2)/2\rfloor} \frac{2k}{2k+1} +
         \sum_{k=1}^{\lfloor(n-1)/2\rfloor} \frac{2k}{2k+1}\right) \\
  & = & \frac{n+1}{8}\left(2\sum_{k=1}^{(n-1)/2}\frac{2k}{2k+1} -
     \frac{n-1}{n}\right) \\
  & = & \frac{n+1}{8}\left(n+H_{\frac{n-1}{2}}-2H_n+\frac{1}{n}\right) ,
\end{eqnarray*}
which again agrees with the claim in the lemma.  The simplified forms
follow the fact that for any odd $n\geq 3$ we can bound
$\frac{1}{n}\leq H_n-H_{\frac{n-1}{2}}\leq \ln 2+\frac{1}{n}$.
\qquad\end{proof}

Combining this result with that of \S\ref{sec:offline}, we see
that our on-line algorithm has expected profit $3/4$ of what could be
obtained with full knowledge of the future.  In terms of competitive
analysis, our algorithm has competitive ratio $4/3$, which means that
not knowing the future is not terribly harmful in this problem!

\subsection{Optimality of Our On-Line Algorithm}

This section proves that algorithm \ouralg\ is optimal.  We will
denote permutations by a small Greek letter with a subscript giving
the size of the permutation; in other words, a permutation on the set
$\{1,2,\ldots,i\}$ may be denoted $\rho_i$ or $\sigma_i$.

A permutation on $i$ items describes fully the first $i$ inputs to our
problem, and given such a permutation we can also compute the
permutation described by the first $i-1$ inputs (or $i-2$, etc.).  We
will use the notation $\sigma_i|_{i-1}$ to denote such a restriction.
This is not just a restriction of the domain of the permutation to
$\{1,\ldots,i-1\}$, since unless $\sigma_i(i)=i$ this simplistic
restriction will not form a valid permutation.

Upon seeing the $i$th input, an algorithm may make one of the
following moves: it may make this input a low selection; it may make
this input a high selection; or it may simply ignore the input and
wait for the next input.  Therefore, any algorithm can be entirely
described by a function which maps permutations (representing inputs
of arbitrary length) into this set of moves.  We denote such a move
function for algorithm $B$ by $M_B$, which for any permutation
$\sigma_i$ maps $M_B(\sigma_i)$ to an element of the set
$\{\mbox{``low''},\mbox{``high''},\mbox{``wait''}\}$.  Notice that not
all move functions give valid algorithms.  For example, it is possible
to define a move function that makes two low selections in a row for
certain inputs, even though this is not allowed by our problem.

We define a generic {\sc holding} state just as we did for our
algorithm. An algorithm is in the {\sc holding} state at time $i$ if
it has made a low selection, but has not yet made a corresponding high
selection.  For algorithm $B$ we define the set $L_B(i)$ to be the set
of permutations on $i$ items that result in the algorithm being in the
{\sc holding} state after processing these $i$ inputs.  We explicitly
define these sets using the move function:
\[ L_B(i) = \left\{ \begin{array}{ll}
       \{\sigma_i | M_B(\sigma_i)=\mbox{``low''} \} & \mbox{ if $i=1$,} \\[1.5ex]
       \{\sigma_i | M_B(\sigma_i)=\mbox{``low''} \mbox{ or } \\
         \hspace*{0.2in} ( M_B(\sigma_i)=\mbox{``wait''} \mbox{ and }
                    \sigma_i|_{i-1}\in L_B(i-1) ) \} & \mbox{ if $i>1$.}
      \end{array}\right. \]
The $L_B(i)$ sets are all we need to compute the expected amortized
profit for interval $i$, since
\begin{eqnarray*}
 E[A_B(i)] & = & Prob[\mbox{Interval $i$ is active}]
               \cdot E[R_{i+1}-R_i|\mbox{Interval $i$ is active}] \\
   & = & \frac{|L_B(i)|}{i!} \left(
               \frac{n+1}{2}-\frac{n+1}{i+1} 
               \sum_{\rho_i\in L_B(i)}\frac{1}{|L_B(i)|} \rho_i(i)\right) \\
   & = & \frac{n+1}{i!}\left(
               \frac{|L_B(i)|}{2} - \frac{1}{i+1}
               \sum_{\rho_i\in L_B(i)}\rho_i(i)\right) .
\end{eqnarray*}
We use the above notation and observations to prove the optimality of
algorithm \ouralg.

\begin{theorem}
Algorithm \ouralg\ is an optimal algorithm for the multiple pair
selection problem.
\end{theorem}

\begin{proof}
Since the move functions (which define specific algorithms)
work on permutations, we will fix an ordering of permutations in order
to compare strategies.  We order permutations first by their size, and
then by a lexicographic ordering of the actual permutations.  When
comparing two different algorithms $B$ and $C$, we start enumerating
permutations in this order and count how many permutations cause the
same move in $B$ and $C$, stopping at the first permutation $\sigma_i$
for which $M_B(\sigma_i)\neq M_C(\sigma_i)$, i.e., the first
permutation for which the algorithms make different moves.  We call
the number of permutations that produce identical moves in this
comparison process the {\em length of agreement between $B$ and $C$}.

To prove the optimality of our algorithm by contradiction, we assume
that it is not optimal, and of all the optimal algorithms let $B$ be
the algorithm with the longest possible length of agreement with our
algorithm \ouralg.  Let $\sigma_k$ be the first permutation in which
$M_B(\sigma_k)\neq M_\ouralg(\sigma_k)$.  Since $B$ is different from
\ouralg\ at this point, at least one of the following cases must hold:

\begin{romannum}
\item[(a)] $\sigma_k|_{k-1}\not\in L_B(k-1)$ and 
           $\sigma_k(k)<\frac{k+1}{2}$ and
           $M_B(\sigma_k)\neq\mbox{``low''}$
(i.e., algorithm $B$ is not in the {\sc holding} state, gets a low
rank input, but does not make it a low selection).
\item[(b)] $\sigma_k|_{k-1}\not\in L_B(k-1)$ and 
           $\sigma_k(k)\geq\frac{k+1}{2}$ and
           $M_B(\sigma_k)\neq\mbox{``wait''}$
(i.e., algorithm $B$ is not in the {\sc holding} state, gets a high
rank input, but makes it a low selection anyway).
\item[(c)] $\sigma_k|_{k-1}\in L_B(k-1)$ and 
           $\sigma_k(k)>\frac{k+1}{2}$ and
           $M_B(\sigma_k)\neq\mbox{``high''}$
(i.e., algorithm $B$ is in the {\sc holding} state, gets a high rank
input, but doesn't make it a high selection).
\item[(d)] $\sigma_k|_{k-1}\in L_B(k-1)$ and 
           $\sigma_k(k)\leq\frac{k+1}{2}$ and
           $M_B(\sigma_k)\neq\mbox{``wait''}$
(i.e., algorithm $B$ is in the {\sc holding} state, gets a low rank
input, but makes it a high selection anyway).
\end{romannum}

In each case, we will show how to transform algorithm
$B$ into a new algorithm $C$ such that $C$ performs at least as well
as $B$, and the length of agreement between $C$ and \ouralg\ is longer
than that between $B$ and \ouralg. This provides the contradiction
that we need.
\begin{description}
\item[{\bf Case (a):}]
Algorithm $C$'s move function is identical to
$B$'s except for the following values:
\[ 
   \begin{array}{l}
 M_C(\sigma_k) = \mbox{``low'',} \\
\ \\
 M_C(\rho_{k+1}) = \left\{ \begin{array}{ll}
         \mbox{``high''} & \mbox{ if $\rho_{k+1}|_k=\sigma_k$ and
                                  $M_B(\sigma_{k+1})=\mbox{``wait''}$ ,} \\
         \mbox{``wait''} & \mbox{ if $\rho_{k+1}|_k=\sigma_k$ and
                                  $M_B(\sigma_{k+1})=\mbox{``low''}$ ,} \\
         M_B(\rho_{k+1}) & \mbox{ otherwise.}
        \end{array}\right.
\end{array} \]
In other words, algorithm $C$ is the same as algorithm $B$ except that
we ``correct $B$'s error'' of not having made this input a low
selection.  The changes of the moves on input $k+1$ insures that
$L_C(k+1)$ is the same as $L_B(k+1)$.
It is easily verified that the new sets $L_C(i)$ (corresponding to the
{\sc holding} state) are identical to the sets $L_B(i)$ for all
$i\neq k$.  The only difference at $k$ is the insertion of
$\sigma_k$, i.e., $L_C(k)=L_B(k)\cup\{\sigma_k\}$.

Let $P_B$ and $P_C$ be the profits of $B$ and $C$, respectively.
Since their amortized costs differ only at interval $k$,
\begin{eqnarray*}
 \hbox to 0pt{$E[P_C-P_B]$\hss} & & \\
    & = & E[A_C(k)] - E[A_B(k)] \\
     & = & \frac{n+1}{k!} \left(
              \frac{|L_C(k)|}{2}-\frac{1}{k+1}
                 \sum_{\rho_k\in L_C(k)}\rho_k(k)\right) \\
     & & \hspace*{0.5in}
- \frac{n+1}{k!}
              \left( \frac{|L_B(k)|}{2}-\frac{1}{k+1}
                 \sum_{\rho_k\in L_B(k)}\rho_k(k)\right) \\
     & = & \frac{n+1}{k!} \left(\frac{1}{2} -
             \frac{1}{k+1}\sigma_k(k)\right) .
\end{eqnarray*}
By one of the conditions of Case (a), $\sigma_k(k)<\frac{k+1}{2}$, so
we finish this derivation by noting that
\[ E[P_C-P_B]  
   = \frac{n+1}{k!} \left(\frac{1}{2} - \frac{1}{k+1}\sigma_k(k)\right)
   > \frac{n+1}{k!} \left(\frac{1}{2} - \frac{1}{k+1}\cdot\frac{k+1}{2}\right)
   = 0 .
\]
Therefore, the expected profit of algorithm $C$ is greater than that
of $B$.

\item[{\bf Case (b):}] As in Case (a) we select a move function for
algorithm $C$ that causes only one change in the sets of {\sc holding}
states, having algorithm $C$ not make input $k$ a low selection.  In
particular, these sets are identical with those of algorithm $B$ with
the one exception that $L_C(k)=L_B(k)-\{\sigma_k\}$.  Analysis similar
to Case (a) shows
\[ E[P_C-P_B] 
   = \frac{n+1}{k!}\left(\frac{1}{k+1}\sigma_k(k)-\frac{1}{2}\right)
   \geq \frac{n+1}{k!}\left(\frac{1}{k+1}\cdot\frac{k+1}{2}-\frac{1}{2}\right)
   = 0 . \]   

\item[{\bf Case (c):}]
In this case we select a move function for algorithm $C$ such that
$L_C(k)=L_B(k)-\{\sigma_k\}$, resulting in algorithm $C$ selecting
input $k$ as a high selection, and giving an expected profit gain of
\[ E[P_C-P_B] 
   = \frac{n+1}{k!}\left(\frac{1}{k+1}\sigma_k(k)-\frac{1}{2}\right)
   > \frac{n+1}{k!}\left(\frac{1}{k+1}\cdot\frac{k+1}{2}-\frac{1}{2}\right)
   = 0 . \]

\item[{\bf Case (d):}]
In this case we select a move function for algorithm $C$ such that
$L_C(k)=L_B(k)\cup\{\sigma_k\}$, resulting in algorithm $C$ not taking
input $k$ as a high selection, and giving an expected profit gain of
\[ E[P_C-P_B] 
   = \frac{n+1}{k!}\left(\frac{1}{2}-\frac{1}{k+1}\sigma_k(k)\right) 
   \geq \frac{n+1}{k!}\left(\frac{1}{2}-\frac{1}{k+1}\cdot\frac{k+1}{2}\right) 
   = 0 . \]

\end{description}

In each case, we transformed algorithm $B$ into a new algorithm $C$
that performs at least as well (and hence must be optimal), and has a
longer length of agreement with algorithm \ouralg\ than $B$ does.
This directly contradicts our selection of $B$ as the optimal
algorithm with the longest length of agreement with \ouralg, and this
contradiction finishes the proof that algorithm \ouralg\ is optimal.
\qquad\end{proof}

\section{Conclusion}
In this paper, we examined a natural on-line problem related to both
financial games and the classic secretary problem.  We select low
and high values from a randomly ordered set of values
presented in an on-line fashion, with the goal of maximizing the
difference in final ranks of such low/high pairs.  We considered two
variations of this problem. The first allowed us to choose only a
single low value followed by a single high value from a sequence of
$n$ values, while the second allowed selection of arbitrarily many
low/high pairs.  We presented provably optimal algorithms for both
variants, gave tight analyses of the performance of these algorithms,
and analyzed how well the on-line performance compares to the optimal
off-line performance.

Our paper opens up many problems. Two particularly interesting directions
are to consider more realistic input sources and to maximize
quantities other than the difference in rank. 

\Appendix

\section{Proof of Expected Final Rank}
In this appendix section, we prove that if an item has relative rank
$x_i$ among the first $i$ inputs, then its expected rank $r_i$ among
all $n$ inputs is given by $E[r_i|x_i]=\frac{n+1}{i+1}x_i$.

\begin{lemma}
If a given item has rank $x$ from among the first $i$ inputs, and if the
$i+1$st input is uniformly distributed over all possible rankings,
then the expected rank of the given item among the first $i+1$ inputs
is $\frac{i+2}{i+1}x$.
\end{lemma}

\begin{proof}
If we let $R$ be a random variable denoting the rank of our given item
from among the first $i+1$ inputs, then we see that the value of $R$
depends on the rank of the $i+1$st input.  In particular, if the rank
of the $i+1$st input is $\leq x$ (which happens with probability
$\frac{x}{i+1}$), then the new rank of our given item will be $x+1$.
On the other hand, if the rank of the $i+1$st input is $>x$ (which
happens with probability $\frac{i+1-x}{i+1}$), then the rank of our
given item is still $x$ among the first $i+1$ inputs.  Using this
observation, we see that
\[  E[R] = \frac{x}{i+1}(x+1) + \frac{i+1-x}{i+1} x
         = \frac{x+1+i+1-x}{i+1} x
         = \frac{i+2}{i+1} x ,
\]
which is what is claimed in the lemma.
\qquad\end{proof}

For a fixed position $i$, the above extension of rank to position
$i+1$ is a constant times the rank of the item among the first
$i$ inputs.  Because of this, we can simply extend this lemma to the
case where $x$ is not a fixed rank but is a random variable, and we
know the expected rank among the first $i$ items.

\begin{corollary}
If a given item has expected rank $x$ from among the first $i$ inputs,
and if the $i+1$st input is uniformly distributed over all possible
rankings, then the expected rank of the given item among the first
$i+1$ inputs is $\frac{i+2}{i+1}x$.
\end{corollary}

Simply multiplying together the change in expected rank from among $i$
inputs, to among $i+1$ inputs, to among $i+2$ inputs, and so on up
to $n$ inputs, we get a telescoping product with cancellations between
successive terms, resulting in the following corollary.

\begin{corollary}
If a given item has rank $x$ from among the first $i$ inputs, and if the
remaining inputs are uniformly distributed over all possible rankings,
then the expected rank of the given item among all $n$ inputs
is $\frac{n+1}{i+1}x$.
\end{corollary}

\bibliographystyle{siam}
\bibliography{rangemax}

\end{document}